%% file: paper.tex
\documentclass[conference]{IEEEtran}
\usepackage{color}
\usepackage{url}
\usepackage{times}
\usepackage{subcaption}
\usepackage{slashbox}
\usepackage{graphicx}
\usepackage{tikz}
\usepackage{amssymb,mathtools}
\usepackage{tikz-3dplot}
\usepackage{epstopdf}
\usepackage[font=bf,aboveskip=2pt]{caption}
\usepackage{amsmath}
\usepackage{algorithm}
\usepackage[noend]{algpseudocode}
\usepackage{soul}
\sethlcolor{green}
\usepackage{breakurl}

\newcommand{\Subsubsection}[1]{\vspace{-0ex} \subsubsection{#1} \vspace{-0ex}}

\newcommand{\eat}[1]{}

	\title{When Edge Meets FaaS: Opportunities and Challenges}

	\author{\IEEEauthorblockN{Runyu Jin\thanks{*}}
 \IEEEauthorblockA{
 \textit{IBM Almaden Research Center}\\
 San Jose, California \\
 runyu.jin@ibm.com \thanks{\thankssymb{*}This work was done when the author was a Ph.D. student at Arizona State University.}
}\and
 \IEEEauthorblockN{Qirui Yang}
 \IEEEauthorblockA{
 \textit{Arizona State University}\\
 Tempe, Arizona \\
 qyang30@asu.edu
 }\and
 \IEEEauthorblockN{Ming Zhao}
 \IEEEauthorblockA{
 \textit{Arizona State University}\\
 Tempe, Arizona \\
 mingzhao@asu.edu
\thanks{Partly supported by National Science Foundation 2126291, 1955593}
 } }
 
\begin{document}
 \maketitle
\IEEEoverridecommandlockouts

	\begin{abstract}
		\input{abstract.tex}
	\end{abstract}
	
	
	
	\input{introduction.tex}
\input{motivation.tex}
	\input{evaluation.tex}

	
	
	
	
	
	
	
	
	
	{\bibliographystyle{plain}
		\bibliography{runyu}}

\end{document}

%% file: abstract.tex
The proliferation of edge devices and the rapid growth of IoT data have called forth the edge computing paradigm. Function-as-a-service (FaaS) is a promising computing paradigm to realize edge computing. This paper explores the feasibility and advantages of FaaS-based edge computing. It also studies the research challenges that should be addressed in the design of such systems, which are 1) the quick decomposing and recomposing of applications, 2) the trade-off between performance and isolation of sandbox mechanisms, and 3) distributed scheduling. The challenges are illustrated by evaluating existing FaaS-based edge platforms, AWS IoT Greengrass, and OpenFaaS.


%% file: introduction.tex
\section{Introduction}
\label{sec:intro}

The proliferation of edge devices and the rapid growth of edge data challenged the traditional cloud computing. In early settings, edge devices produced data and sent it to cloud for computation and storage. The high costs of cloud services and the network latency caused by data transportation outweigh the high-performance cloud can provide. Edge computing emerged to solve the problem. In edge computing, edge devices and edge servers are considered as important resources that can help with computation and data storage; data generated on edge devices is processed locally to avoid network latency. This can greatly reduce application response time which is especially important for time-sensitive edge applications. Also, many edge devices generate data concerning user privacy. The data is more secure to be stored on local devices than on a shared cloud.

A proper abstraction is key to enabling computing across the heterogeneous resources on the edge and across the multiple tiers of resources from the edge to the cloud. In this paper, we argue that Function-as-a-Service (FaaS), an emerging cloud computing model can provide the much needed abstraction for edge computing. In FaaS, the unit of computation is a function. When a request is received, the platform starts an ephemeral execution sandbox for the function. When load increases, it quickly and dynamically increases the number of execution units. As soon as the function finishes, the sandboxes are terminated. 

To investigate the opportunities and challenges, we considered the existing commercial FaaS platforms for edge computing such as AWS IoT Greengrass~\cite{greengrass}. It enables the edge workloads to be computed locally using AWS Lambda~\cite{awslambda}. 
We also developed our own prototype based on OpenFaaS~\cite{openfaas}, which is an open-source FaaS platform that supports edge devices. Using these platforms we study the challenges of FaaS for edge computing. We conducted the experiments on local edge devices and a edge server.

%% file: motivation.tex
\section{Motivations}
\label{sec:motivation}

\eat{Section 3: Motivations
3.1: What are the potential benefits of serverless edge computing? (explain the benefits; explain the limitations of existing edge computing paradigms)
3.2: what are the unknown questions to serverless edge computing? (explain the questions and their importance)}



We advocate FaaS as the abstraction for edge computing because it can greatly help reap the benefits of edge computing and address its challenges by providing the following.

\noindent\textbf{Faster Responses:} 
First, FaaS improves edge applications' response time by doing more computation closer to data sources. 
Compared to complex applications, functions are small in size and resource-conserving. They can better fit into the limited resources on edge devices. 

Second, FaaS can better exploit the heterogeneous resources available at the edge to deliver faster responses.
Applications often contain functions that have different workloads. Functions with I/O intensive workloads can gain optimal performance running on flash storage. Functions with computation intensive workloads can boost the performance using hardware accelerators such as GPUs and FPGAs. 

Third, FaaS can reuse functions for different applications to save function initialization time. Various applications contain the same function. 
For example, many personal virtual assistant applications contain the text-to-speech function to interact with users. The frequently used functions can be reused by different applications to avoid the cold start. 


\noindent\textbf{Better Privacy:} FaaS isolates each function to its own address space. A function generally involves less user data compared to an application. This facilitates better privacy since FaaS prevents data from being leaked or modified at a large scale on edge devices.

\noindent\textbf{Higher Productivity:} 
Edge devices are highly heterogeneous with diverse software and hardware. This prevents the utilization of distributed edge resources. FaaS provides a nice abstraction to hide the heterogeneity and enables the productive use of the diverse edge resources.  

\noindent\textbf{More Reliability:}
One key benefit of FaaS is to provide more reliability by employing function-based sandboxes. On edge devices, due to the continuous interaction with the physical world, they are more prone to failures that can bring severe consequences (think of a camera failure on an autonomous driving car). FaaS restraints the failure within a function sandbox to improve the reliability. 

\noindent\textbf{Lower Cost:}
FaaS can lower the cost of users by utilizing more edge resources. As discussed above, FaaS can better fit functions in edge resources to reduce the demand of cloud servers. More requests are served locally and the cost is reduced.



\input{challenges.tex}

%% file: challenges.tex
\section{Challenges}
\label{sec:problems}

To deliver effective FaaS-based edge computing, we identify the following key challenges.

\noindent\textbf{From Applications to Functions: }
As illustrated in Section~\ref{sec:motivation}, FaaS edge computing benefits largely from the finer-grained execution unit. However, benefits arise with challenges. Many functions need to be chained to work in a pipelined manner according to the application logic. The overhead of executing sequential functions to get final results can be high if the system is not cautiously designed. In Section~\ref{sec:eval}, we show that function chaining overhead is not significant for functions deployed within an edge device. However, when the functions are distributed across devices or across edge and cloud, both computation and communication overhead needs to be carefully considered.

\noindent\textbf{Sandbox Mechanisms: } 
Sandbox mechanisms is a key component of FaaS, which includes virtual machines (VM), containers~\cite{docker, gvisor} and other lightweight virtualization systems~\cite{firecracker} that isolate functions from each other to provide performance and security guarantees. 
While providing necessary isolation, sandboxes unavoidably add extra overhead to run functions. 

There are two aspects to look at the sandbox mechanisms: performance and isolation. Performance represents how fast a function runs and responds. Isolation refers to performance isolation which maintains performance when other functions run together with the functions inside the sandbox, failure isolation which restraints the failure within the sandbox, and security isolation that secures the data of a sandbox. 
In edge computing, time-sensitive edge applications demand both the small performance overhead and strong performance isolation. The balance point of the two aspects needs to be figured out to achieve the best trade off for the target application and scenario. We evaluate existing popular sandboxes for edge computing in Section~\ref{sec:eval} and study the trade off between performance and isolation.

\noindent\textbf{Distributed Scheduling: }
FaaS-based edge computing increases the scheduling flexibility by providing smaller scheduling units, which leads to the complexity of distributed scheduling. 
To achieve distributed scheduling, at least two factors need to be considered: scheduling horizontally within edge devices and scheduling vertically across the edge and cloud.
The various constraints that scheduling must consider for resources (e.g., power/battery capacity, availability) and applications (e.g., privacy) on the edge further adds to its complexity.

The challenges in horizontal scheduling is in how to efficiently utilize the edge resources. Applying FaaS decouples the application logic and the hardware which executes the program. How to effectively map functions to edge devices equipped with heterogeneous hardware resources to accelerate applications remains a question. 

The challenges in vertical scheduling is in how to fully utilize every tier bottom up from the edge to the cloud. Given that devices on the path from the edge to the cloud are usually less powerful than cloud servers, scheduling functions vertically to the cloud can speedup the response time for some functions. 

%% file: evaluation.tex
\vspace{-10pt}
\section{Evaluation}
\label{sec:eval}

\vspace{-5pt}


\subsection{Sandbox Mechanisms}

AWS Greengrass supports three sandbox mechanisms to isolate the function executions which are Greengrass container (GGC), a lightweight container that makes use of cgroups and namespaces for isolation, docker container (DOCK) and Greengrass no container (GNC). We first evaluate the run time overhead of each sandbox mechanism. We used a python function, image recognition, as an example for evaluation. It recognizes objects in the image using the SqueezeNet deep learning model~\cite{IandolaMAHDK16} executed on top of the MXNet framework~\cite{mxnet}. We recorded the timestamps of sending the function trigger and the start and end of the computation to get the function's run time and compute time. We used the default MQTT topic pub/sub system for function triggers. The experiments were done on a Raspberry Pi 3B+ single board computer which is equipped with 4 cores and 1GB memory. 

\begin{figure}[t]
\centering
\captionbox{Performance overhead of different sandbox mechanisms\label{fig:performance_overhead}}{\includegraphics[width=0.5\textwidth, scale=0.5]{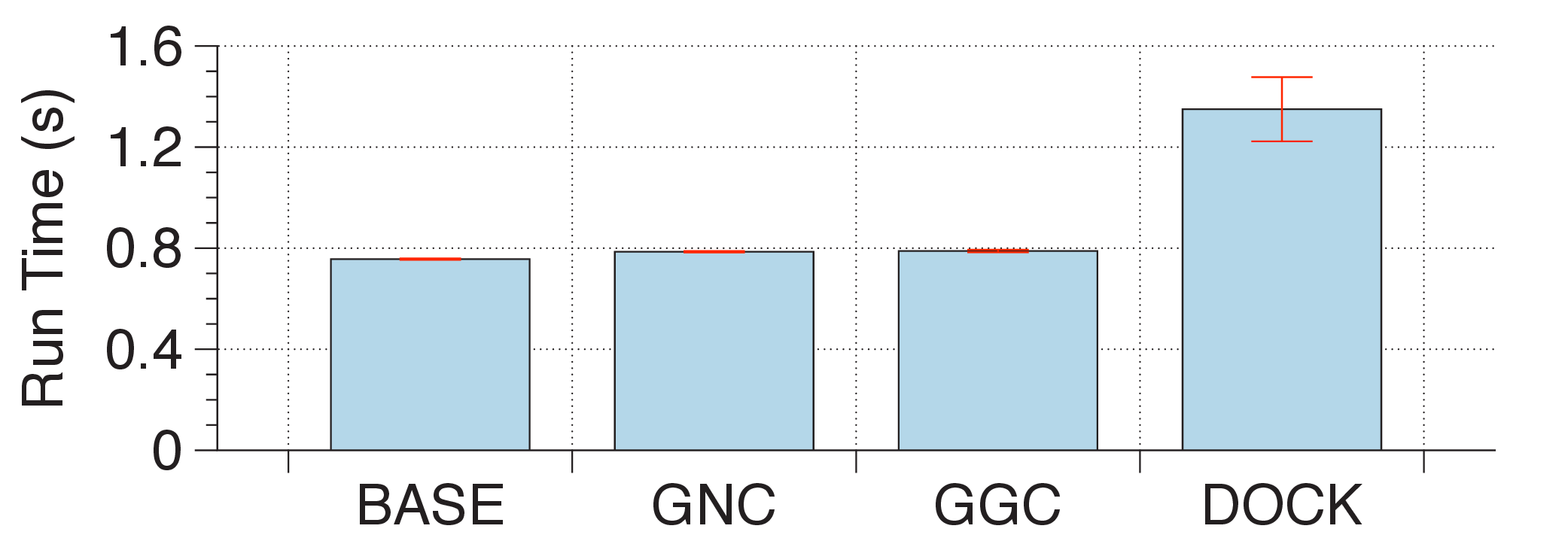}}%
\end{figure}

Figure~\ref{fig:performance_overhead} shows the results. The error bars describe the standard deviation. We compare the sandbox results with the result of the same python code running on the bare metal (BASE). We can see for GNC and GGC, the run time overhead is quite small, at 3.8\% and 4.2\%, respectively. This is because these two sandboxes are light weight with limited isolation. 

\begin{figure}[t]
\centering
\captionbox{Performance isolation of different sandbox mechanisms\label{fig:performance_isolation}}{\includegraphics[width=0.5\textwidth, scale=0.5]{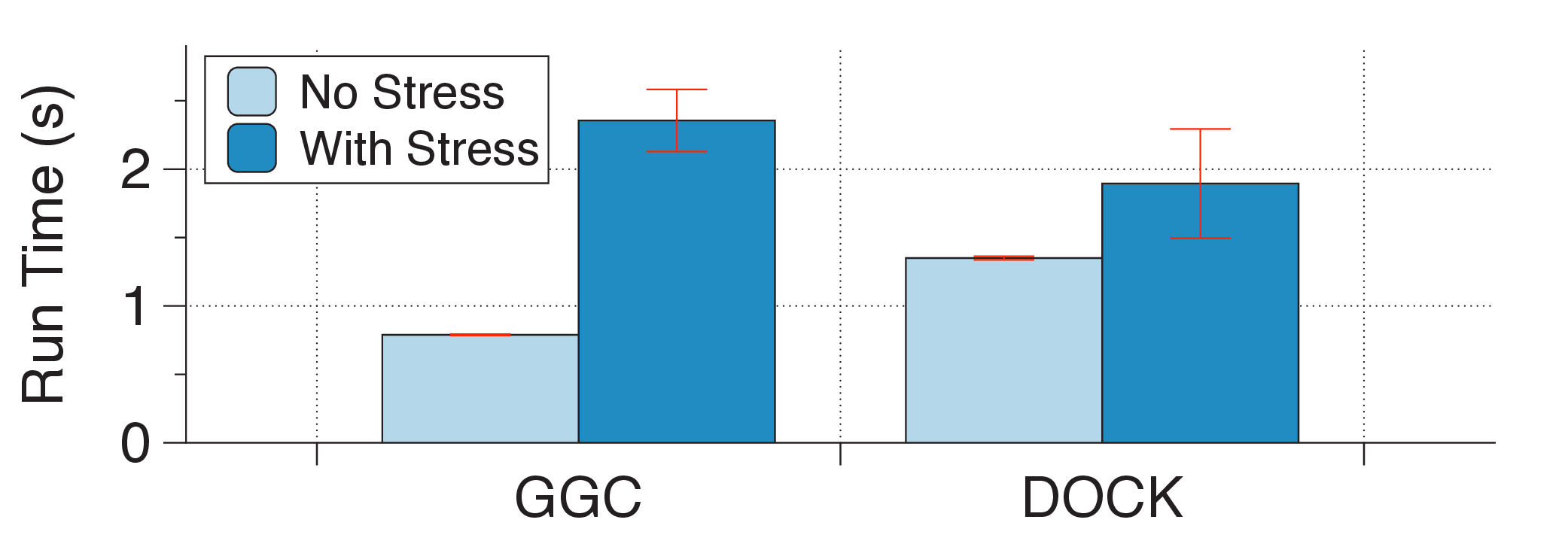}}%
\end{figure}
However, We found that GNC and GCC do not provide good isolation. For GNC, functions can read and write to any files that belong to the user as long as the user's uid and gid is provided to AWS Greengrass. For GGC, functions cannot write to local files but can still read any files belonging to this user. 
For DOCK, which is heavier weight than GGC and GNC, has an overhead of 78.3\%. However, it is more secure than the previous two sandboxes since functions do not have access to host files. 

We then evaluate the isolation of sandbox mechanisms. We use a clone of the function which runs within a busy loop as the stress test. We compare the run time of running the function alone to running the function together with the stress test. Figure~\ref{fig:performance_isolation} shows the result. From the result, we can see that GGC fails to isolate the interference. The response time is slowed down twice that of running the function alone. For DOCK, which has a better isolation mechanism, the overhead is much smaller at 40\%. However, the interference is still big and for some time-sensitive functions, the overhead is unacceptable. 

\begin{figure}[t]
\centering
\captionbox{Function chaining and cold start overhead\label{fig:function_chaining}}{\includegraphics[width=0.5\textwidth, scale=1]{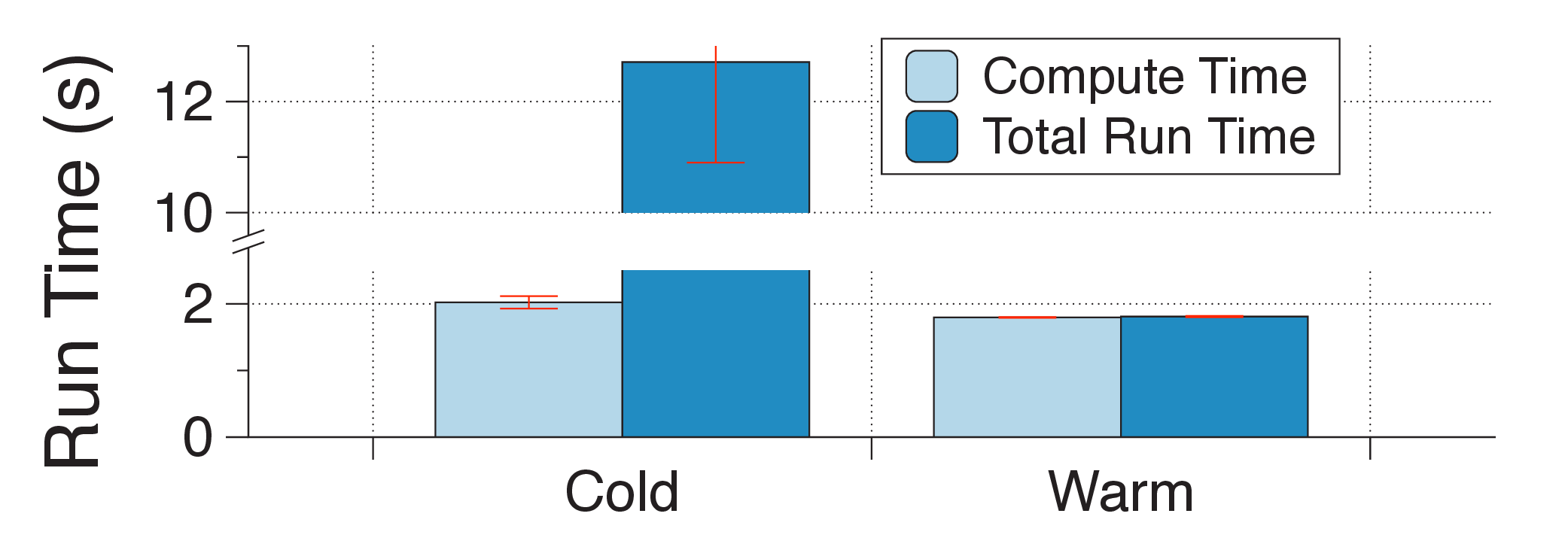}}%
\end{figure}

Finally, we evaluate the cold start overhead of functions and the communication overhead of function chaining. We sequentially ran two functions using the default AWS Greengrass container. The first function uses the on-board camera to take a picture, stores it to the local storage and triggers the second function, which then loads the picture and conducts image recognition. We ran the functions using both cold containers and warm containers and compared the total run time of the two functions to the sum of two functions' compute time. We used the default MQTT topic pub/sub system for function communications. From Figure~\ref{fig:function_chaining}, we can see that for cold containers, the total run time is significantly more (5.3 times) than the actual time required for function executions. The cold start of containers has a huge impact on the application's run time which needs to be further optimized. For warm containers, the overhead is negligible. But we cannot say the current mechanism used for function communication is good considering the weak isolation that Greengrass container provides.

\begin{figure}[t]
\captionbox{How scheduling affects performance\label{fig:scheduling}}{\includegraphics[width=0.5\textwidth, scale=0.5]{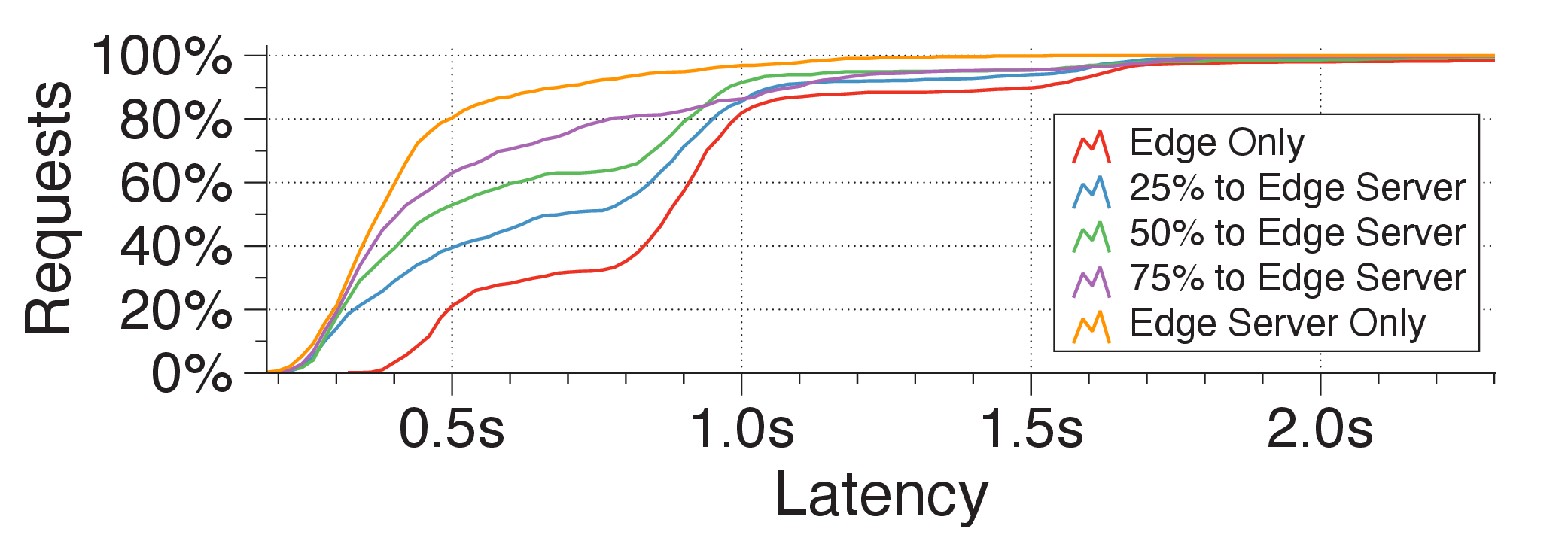}}%
\end{figure}
\subsection{Distributed Scheduling}

We implemented a prototype on OpenFaaS to evaluate distributed scheduling. 
The evaluation involves three platforms: edge only platform, edge and cloud cooperate platform, and cloud only platform. For edge only platform, the scheduling is provided by OpenFaaS which dynamically scales the number of functions to all the local devices when load increases. For edge and cloud together platform, we modified OpenFaaS to offload the requests to a server following a preset proportion. For cloud only platform, to make a fair comparison, we also used OpenFaaS on the server instead of a commercial FaaS platform and offloaded all the requests to the edge server. During the experiment, each edge device issued 3 requests per second for 30 seconds. Each request contains a sentence of 20 words to be transferred to audio speech. The maximum concurrent running functions are set to be 12 and 6 for the edge cluster and server, respectively. We recorded the run time of each request. 


Figure~\ref{fig:scheduling} shows the CDF results of function run time. We set the proportion of offloading to cloud at 25\%, 50\%, and 75\%, respectively. Edge only platform has an average run time latency of 0.86s, whereas a cloud only platform on average takes 0.44s. With the increased amount of offloaded requests, the run time latency decreases. In this example, the naive scheduling policy is by no means showing the best performance of distributed scheduling but it can still confirm the benefits of vertical scaling to nearby servers. In the meanwhile, the reduced performance saves the costs of the cloud service. With a fully-functional scheduling policy, we believe the performance can be further improved and the costs can be further reduced.